\documentclass[twocolumn,showpacs,preprintnumbers,prl,aps,superscriptaddress]{revtex4-2}
\usepackage{graphicx}
\usepackage{amsmath}
\usepackage{amssymb}
\usepackage[retain-unity-mantissa=false]{siunitx}
\usepackage{graphicx}
\usepackage{xcolor}
\usepackage[caption=false]{subfig}
\usepackage{placeins}
\usepackage[colorlinks=true, linkcolor=blue, citecolor=blue, filecolor=blue, urlcolor=blue,pdfproducer={.},pdfcreator={.}]{hyperref}
\usepackage{cleveref}
\usepackage{blindtext}
\usepackage{comment}
\usepackage{verbatim}
\usepackage{ulem}
\usepackage[utf8]{inputenc}
\usepackage[T1]{fontenc} 
\usepackage{lmodern} 
\usepackage{physics}
\usepackage{booktabs} 
\usepackage{ragged2e} 
\usepackage{array} 

\newcommand{\mutm}{$\mu$TM }
\newcommand{\mutml}{$\mu$T-model }
\newcommand{\supp}{Supplementary Information}

\renewcommand{\eqref}[1]{equation (\ref{#1})}

\begin{document}
    \title{Control of transport phenomena in magnetic heterostructures by wavelength modulation}
    \author{Christopher~\surname{Seibel}}
    \email{cseibel@physik.uni-kl.de}
    \author{Marius~Weber}
    \author{Martin~Stiehl}
    \author{Sebastian~T.~Weber}
    \author{Martin~Aeschlimann}
    \author{Hans~Christian~Schneider}
    \affiliation{Department of Physics and Research Center OPTIMAS, Technische Universitaet Kaiserslautern, 67663 Kaiserslautern, Germany}
    \author{Benjamin~Stadtmüller}
    \affiliation{Department of Physics and Research Center OPTIMAS, Technische Universitaet Kaiserslautern, 67663 Kaiserslautern, Germany}
    \affiliation{Institute of Physics, Johannes Gutenberg University Mainz, Staudingerweg 7, 55128 Mainz, Germany}
    \author{Baerbel~Rethfeld}
    \affiliation{Department of Physics and Research Center OPTIMAS, Technische Universitaet Kaiserslautern, 67663 Kaiserslautern, Germany}

    \date{\today}
\begin{abstract}
We demonstrate the tuneablity of the ultrafast energy flow in magnetic/non-magnetic bilayer structures by changing the wavelength of the optical excitation. This is achieved by an advanced description of the temperature based \mutml that explicitly considers the wavelength- and layer-dependent absorption profile within multilayer structures. For the exemplary case of a Ni/Au bilayer, our simulations predict that the energy flow from Ni to Au is reversed when changing the wavelength of the excitation from the infrared to the ultraviolet spectral range. These predictions are fully supported by characteristic signatures in the magneto-optical Kerr traces of the Ni/Au model system. Our results will open up new avenues to steer and control the energy transport in designed magnetic multilayer for ultrafast spintronic applications. 
 \end{abstract}
    \maketitle


The increasing demand for storing and processing digital information with enhanced speed and energy efficiency has triggered the search for new concepts to control binary information in condensed matter systems. The most direct way for pushing information technology to higher frequencies is to employ ultrashort light pulses to manipulate the spin degree of freedom in spintronic device structures. The foundation for this approach was laid by pioneering studies~\cite{Beaurepaire1996,Koopmans2010,Lalieu2019,Kimel2019,Battiato2010,Bigot2009} demonstrating the optical manipulation of individual ultrathin magnetic layers on ultrafast, sub-picosecond timescales.    

In more realistic spintronic multilayer structures, the magnetization dynamics is not only governed by local spin flip scattering processes within the individual layers \cite{Mueller2013PRL,Cinchetti2006}. 
It is also strongly influenced by energy and (spin-dependent) particle transport between the individual layers. For instance, spin-dependent transport can strongly increase the speed of the demagnetization process~\cite{Battiato2010, Rudolf2012, Hofherr2017,Eschenlohr2013,Melnikov2011} of a magnetic layer or can even alter the magnetic order of a collector layer~\cite{Schellekens2013}. In a similar way, energy transport can alter the recovery process of the magnetic order (remagnetization) after the optical excitation~\cite{Pudell2018,Kazantseva2007,Bierbrauer2017}.  It is therefore of utmost importance to devise new concepts to steer and control the strength and direction of the energy and particle transport in complex multilayer systems.

From a fundamental point of view, energy and particle transport in multilayers are directly linked to the spatial absorption profile of the  exciting light field in the multilayer structure. 
The resulting
gradients
of
temperature and chemical potentials 
across the interfaces
are ultimately responsible for energy and (spin-dependent) particle transport between adjacent layers and thus determine the ultrafast magnetization dynamics within multilayer structures~\cite{Rouzegar2021arxiv,Melnikov2011}. 

So far, only a few experimental studies have reported characteristic changes of the ultrafast demagnetization dynamics of multilayer stacks for different layer-dependent absorption profiles~\cite{Eschenlohr2013,Pudell2018,Cardin2020},
which have been
realized by altering the wavelength of the optical excitation. In this way, Cardin et al. demonstrated a correlation between the magnitude of the loss of magnetic order in a Co/Pt multilayer structure and the spatial extension of the electromagnetic energy deposited into the material system~\cite{Cardin2020}. On the other hand, Pudell et al. uncovered a rapid energy transfer in a magnetic/non-magnetic bilayer system that results in an almost identical magnetization dynamics independent on the spatial excitation profile within the bilayer structure~\cite{Pudell2018}. Despite these intriguing experimental observations, there is no clear theoretical approach to control the strength and flow direction of these transport processes between adjacent layers of a multilayer structure.

In this manuscript, we build on these intriguing experimental observations and demonstrate that the direction of the energy and heat flow in magnetic/non-magnetic multilayer structures can be controlled by the wavelength of the optical excitation. Our conclusions are based on a considerable extension of the temperature-based \mutml\cite{Mueller2014PRB} that explicitly considers the wavelength- and layer-dependent absorption profile as well as the energy transport and spin-dependent particle transport within a multilayer structure. 
The predictive power of our model simulations is confirmed by time-resolved magneto-optical Kerr studies of the ultrafast magnetization dynamics in a Ni/Au bilayer structure. Our findings will open up the way towards engineering and controlling energy and particle transfer phenomena in designed multilayer structures for the next generation of spintronic applications operating on sub-picosecond timescales. 

\begin{figure}
    \centering
    \includegraphics[width=\columnwidth]{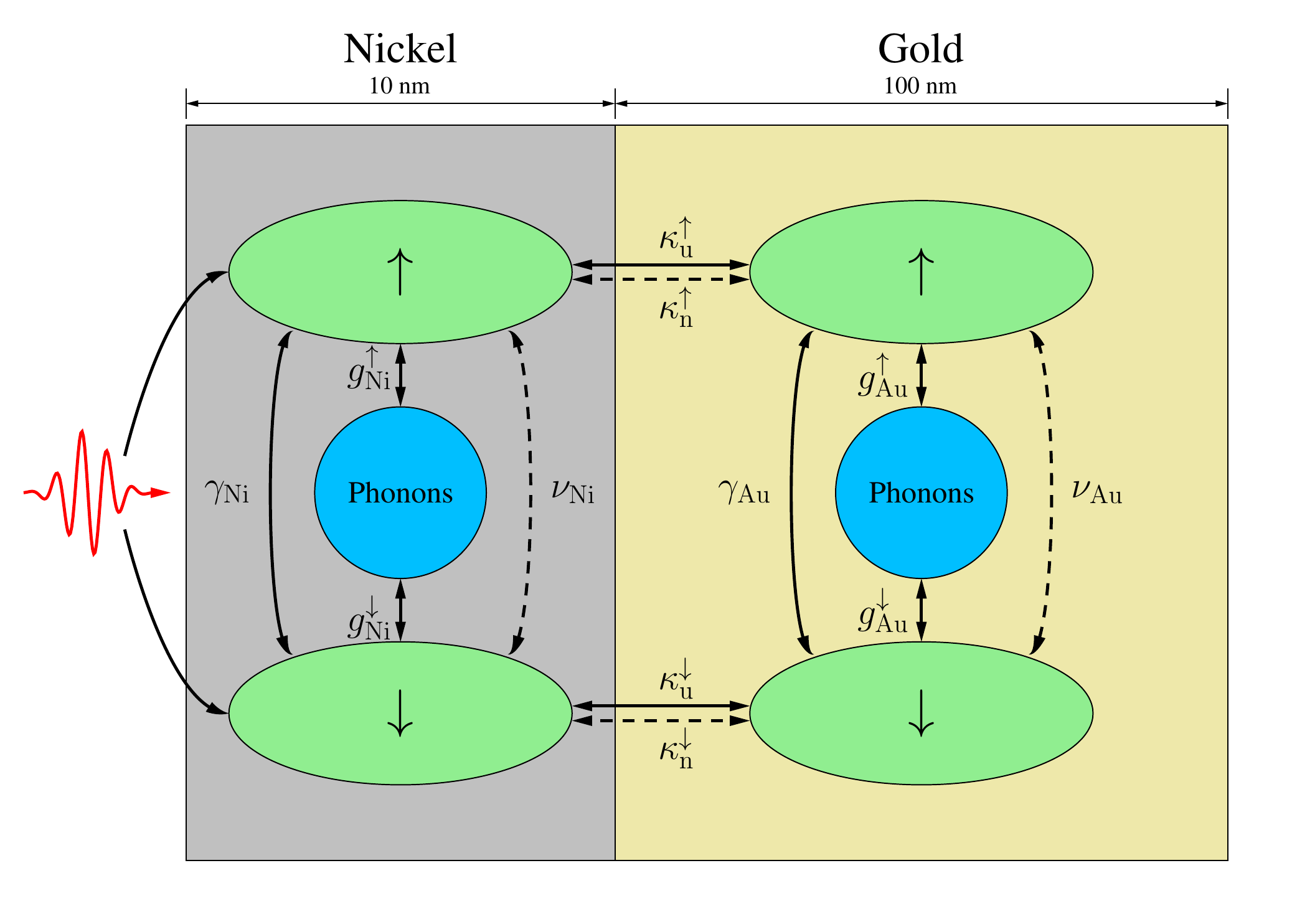}
    \caption{Interaction scheme of Ni on top of Au as substrate. Solid lines indicate energy transfer and dashed lines particle transfer. The transport and coupling parameters are defined in \eqref{eq:mutm_bilayer}. \label{fig:ni_au_scheme}}
\end{figure}

In order to simulate the magnetization dynamics of the optically excited ferromagnet in the bilayer system, we apply the temperature-based \mutml ($\mu$TM) \cite{Mueller2014PRB}. 
It extends the well-known two-temperature model (TTM)~\cite{Anisimov1974} by separating the temperatures of spin-up and spin-down electrons 
and additionally traces the chemical potentials 
of both spin directions as illustrated in \autoref{fig:ni_au_scheme}.
Kinetic calculations strongly suggest the equilibration of the chemical potentials after distortion by laser excitation as the driving force for ultrafast magnetization dynamics~\cite{Mueller2011}, and this effect is captured in the $\mu$TM.
For the description of a bilayer, the \mutm is well suited, because it can be extended to particle and energy transport across interfaces.
The \mutm then consists of an equation system that describes the changes of the energy density $u$ of spin-resolved electrons and phonons and the particle densities $n$ of the electronic subsystems due to different equilibration processes between all subsystems. It reads 
\begin{subequations} \label{eq:mutm_bilayer}
\begin{align}
	\begin{split}
		\dv{u_M^\sigma}{t}   &= -\gamma_M ( T_M^\sigma - T_M^{\overline{\sigma}}) - g_M^\sigma (T_M^\sigma - T_M^p) + s_M^\sigma(t)\\
 		&~~~+ \mathrm{max}\left(\mu_M^\uparrow, \mu_M^\downarrow\right)\nu_M (\mu_M^\sigma - \mu_M^{\overline{\sigma}})\\
        &~~~-\frac{\kappa_{u,T}^\sigma}{d_M}(T_M^\sigma - T_{\overline{M}}^\sigma) - \frac{\kappa_{u,\mu}^\sigma}{d_M}(\mu_M^\sigma - \mu_{\overline{M}}^\sigma), \label{eq:muTM_bilayer_dudt}
 	\end{split}\\
    \dv{u_M^p}{t}&= -g_M^\uparrow (T_M^p - T_M^\uparrow) - g_M^\downarrow (T_M^p - T_M^\downarrow), \label{eq:muTM_bilayer_phonons}\\
    \dv{n_M^\sigma}{t}   &= -\nu_M (\mu_M^\sigma - \mu_M^{\overline{\sigma}}) - \frac{1}{d_M}j_M^\sigma \enspace, \label{eq:muTM_bilayer_dndt}
\end{align}
\end{subequations}
where the superscript $\sigma \in \{\uparrow, \downarrow \}$ denotes the spin direction of an electronic subsystem, $\overline{\sigma}$ the opposite direction, and $p$ the phonons. 
Quantities labeled by $M$ refer to the magnetic material, whereas $\overline{M}$ denotes the non-magnetic layer of the bilayer. 
The optical excitation  of the electronic subsystems is modelled by the laser source term $s_M^\sigma(t)$.
The equilibration processes of temperatures $T$  and chemical potentials $\mu$ within one layer are driven
 by an exchange 
energy and particles between the subsystems.
Similar to the conventional $\mu$TM, they are determined by the electron-phonon coupling parameter $g$ and the energy- and particle coupling between up and down electrons, $\gamma$ and $\nu$, respectively~\cite{Mueller2014PRB}. 
The transport across the interface is described by the terms containing the transport parameters $\kappa$. 
Transport of quantity $a$ due to a gradient or difference of quantity $b$ is determined by the corresponding transport parameter $\kappa_{a,b}$.
All transport parameters can be derived theoretically, which is shown in the \supp. 
The particle transport appears in  \eqref{eq:muTM_bilayer_dndt} as current $j_M^\sigma$, which consists of spin- and charge currents. 
The spin-polarized current $j_{S,\mathrm{M}}^\sigma$ being injected from the magnetic into the non-magnetic layer is defined as
\begin{equation}
    j_{S,\mathrm{M}}^\sigma = (1- R_\mathrm{M}^\sigma ) \left[\kappa_{n,T}^\sigma(T_\mathrm{M}^\sigma - T_\mathrm{\overline{M}}^\sigma) + \kappa_{n,\mu}^\sigma(\mu_\mathrm{M}^\sigma - \mu_\mathrm{\overline{M}}^\sigma)\right], \label{eq:spin_current_fm}
\end{equation}
applying spin- and material-dependent interface reflectivities $R_M^\sigma$ based on first-principles to account for the partial reflection of the currents at the interface~\cite{Lu2020}.
To conserve charge neutrality, a current of the same amount has to flow back into the magnetic layer. This charge current is given by
\begin{equation}
    j_{C,\mathrm{M}}^\sigma = \frac{1-R_\mathrm{\overline {M}}^\sigma}{2-R_\mathrm{\overline{M}}^\uparrow - R_\mathrm{\overline{M}}^\downarrow}\left[(1-R_\mathrm{M}^\uparrow)j_{S,\mathrm{M}}^\uparrow + (1-R_\mathrm{M}^\downarrow)j_{S,\mathrm{M}}^\downarrow \right]. \label{eq:charge_current_fm}
\end{equation}
Together, we denote the total interface current for the magnetic material by $j_\mathrm{M}^\sigma = j_{S,\mathrm{M}}^\sigma - j_{C,\mathrm{M}}^\sigma$, and $j_\mathrm{\overline{M}}^\sigma = - j_\mathrm{M}^\sigma$ for the non-magnetic layer.
We assume that energy and particles in both layers are distributed homogeneously over the respective material. 
This is justified for thicknesses smaller than the ballistic range, which for the here-considered non-magnetic layer of gold is about \SI{100}{\nm}~\cite{Hohlfeld1997, Brorson1987}.

\begin{figure}
    \centering
    \includegraphics[width=\columnwidth]{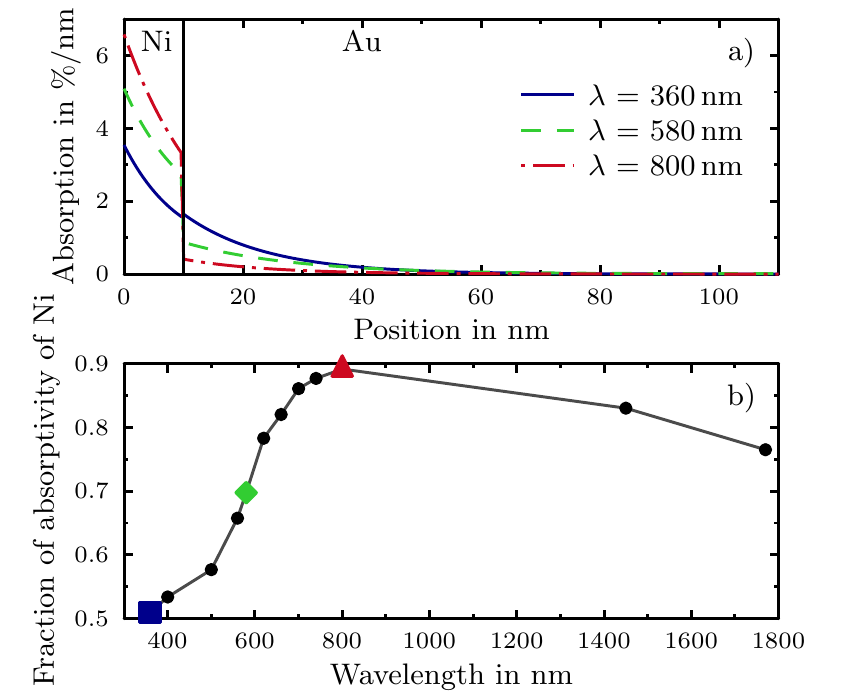}
    \caption{a) Calculated absorption profiles of a Ni[\SI{10}{\nano\meter}]\-|Au[\SI{100}{\nano\meter}]\-|MgO[\SI{500}{\nm}] heterostructure for the three wavelengths marked in b). b) Ratio between the absorptivity of Ni and the total absorptivity of the Ni|Au bilayer for various wavelengths. In the visible range, a larger fraction of energy is absorbed in Ni for longer wavelengths. The line is a guide to the eye.\label{fig:absorption}}
\end{figure}

One of the key ingredients of this study is the wavelength- and layer-dependent absorption profile of the laser light in the bilayer system. It determines the energy content $s_M(t)$, i.e., the strength of the optical excitation, in each individual layer. 
To that end, we numerically solve the Helmholtz equation for a sample consisting of a \SI{10}{nm} Nickel layer on a \SI{100}{\nm} gold film, grown on an insulating substrate (MgO, \SI{500}{\nm} thickness)~\footnote{See \supp~for details on the calculation of the absorption profiles}.
The refractive indices entering the Helmholtz equation have been obtained with density functional theory (DFT) calculations for Au and Ni~\cite{Werner2009} and from experiments for MgO~\cite{Stephens1952}. 
\autoref{fig:absorption}~a) shows the calculated absorption profiles in nickel and gold for three different wavelengths in the visible range. 
The energy distribution within the bilayer 
strongly depends on the applied laser wavelength. 

To determine the deposited energy in the individual layers, we integrate the depth-dependent absorptivity $\mathrm{d}A(z)$ in the individual layers. This yields the total absorbed energy proportional to $A =
\int_0^d \dv{A(z)}{z} \mathrm{d}z$.
\autoref{fig:absorption}~b) shows the fraction of the integrated absorptivity 
in the nickel layer, $A_\mathrm{Ni}$,
normalized
to the total absorption of the bilayer,
$A_\mathrm{Ni} + A_\mathrm{Au} $. 
At small wavelengths in the ultraviolet regime, only  \SI{50}{\percent} of the light's energy is absorbed in the Nickel layer. This fraction increases up to almost \SI{90}{\percent} for \SI{800}{\nm}, i.e., for the wavelength most frequently employed in ultrafast magnetization dynamics studies.
Beyond that, the absorption decreases again for even larger wavelength in the infrared regime. 
This layer- and wavelength-dependent light absorption allows us to formulate a much more realistic 
 description of the optical excitation processes in magnetic multilayer structures. 
Considering a Gaussian laser pulse $I(t)$ and an equal energy absorption of minority and majority electrons~\footnote{Note that for the observables studied in this work, 
the influence of the distribution of absorbed energy to the up and down electrons is negligible, see 
\supp. We therefore choose an equal distribution.}, the source term entering \eqref{eq:muTM_bilayer_dudt} can be expressed as 
\begin{equation}
    s_M^\sigma (t) = \frac{A_M}{2d} I(t) \enspace. \label{eq:source_term}
\end{equation}


This source term allows us now to calculate the 
interface-gradients of the temperature and chemical potential of a bilayer system.
To this end, we solve the coupled differential equations (\ref{eq:mutm_bilayer}) of the \mutml numerically applying the Crank-Nicolson method~\cite{CrankNicolson1947}.
Initially, the system is at room temperature. 
All parameters of the calculations are summarized in the \supp.
temperatures and the chemical potentials are determined at each instant in time by the transient energy densities $u$ and particle densities $n$. They can be extracted by a root-finding method, evaluating the 0th and 1st moment of the corresponding Fermi distributions with DFT-calculated densities of states (DOS)~\cite{Lin2008}. 

\begin{figure}
    \centering
    \includegraphics[width=\columnwidth]{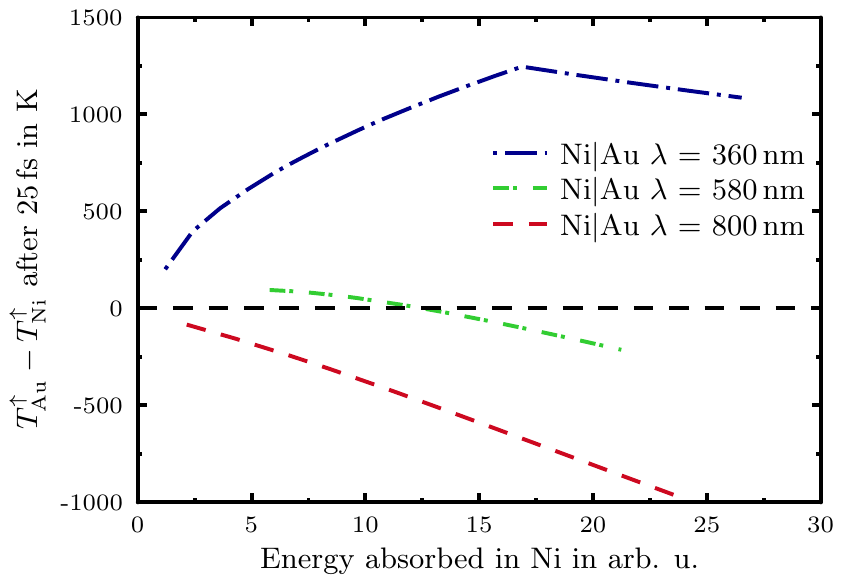}
    \caption{Difference between spin up temperatures of Au and Ni in dependence of the energy absorbed in Ni at $t=\SI{25}{\femto\second}$. This represents the energy flow between the two materials. Negative signs indicate an energy flow from Ni to Au whereas the opposite holds for positive signs. For \SI{360}{\nm}, an additional back-heating of the Ni layer by the Au layer is clearly present.\label{fig:temperature_difference}}
\end{figure}
\autoref{fig:temperature_difference} shows the temperature difference between the Ni and Au layer right after the optical excitation 
for three characteristic photon energies in dependence on the 
energy absorbed in the Ni layer. This temperature gradient is directly responsible for the energy transfer between the layers. Overall, we find striking differences in the temperature gradients 
depending on the wavelength of the optical excitation. 
For \SI{360}{\nm}, the temperature of the Au layer exceeds the one in Ni for all excitation strengths despite the almost identical energy absorption in both layers (see Fig.~2b). This temperature gradient favors an energy flow from Au to Ni and hence leads to a counterintuitive heating of Ni by the Au layer. In contrast, we find a larger temperature of the Ni layer for all excitation strengths with \SI{800}{\nm} photons pointing to an energy transport from the Ni into the Au layer. For the intermediate wavelength of \SI{580}{\nm}, the temperatures of the Ni and Au layers are almost identical.

The observed wavelength-dependent temperature gradient across the magnetic bilayer system can be attributed to two major ingredients: (i) the layer dependent absorption profile within the bilayer structure and (ii) the specific electronic heat capacity of the individual layers. The heat capacity is a material parameter and hence independent of the wavelength of the optical excitation. Consequently, the sign and magnitude of the temperature gradient and the corresponding energy transport within a magnetic bilayer structure is indeed solely determined by the wavelength-dependent absorption within the individual layers. In this way, our model clearly demonstrates the possibility to tune and control of the sign and magnitude of the interlayer energy transport by changing the excitation wavelength. 
At present, we cannot verify our predictions by directly monitoring the temperature gradients within a bilayer systems experimentally, as these quantities are extremely challenging to access. Instead, we focus on characteristic signatures of the wavelength-dependent energy transport in the ultrafast magnetization traces of optically excited bilayer structures.  

\begin{figure}
    \includegraphics[width=\columnwidth]{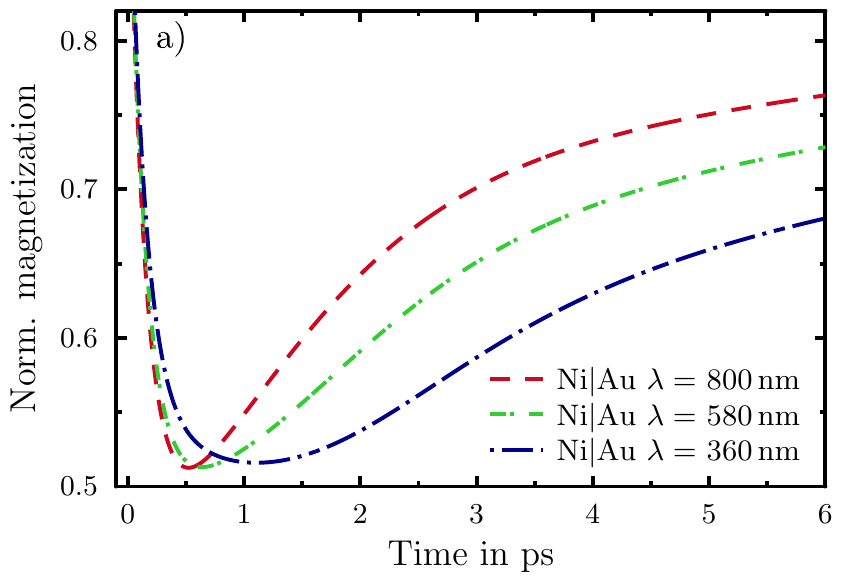}\\

    \includegraphics[width=\columnwidth]{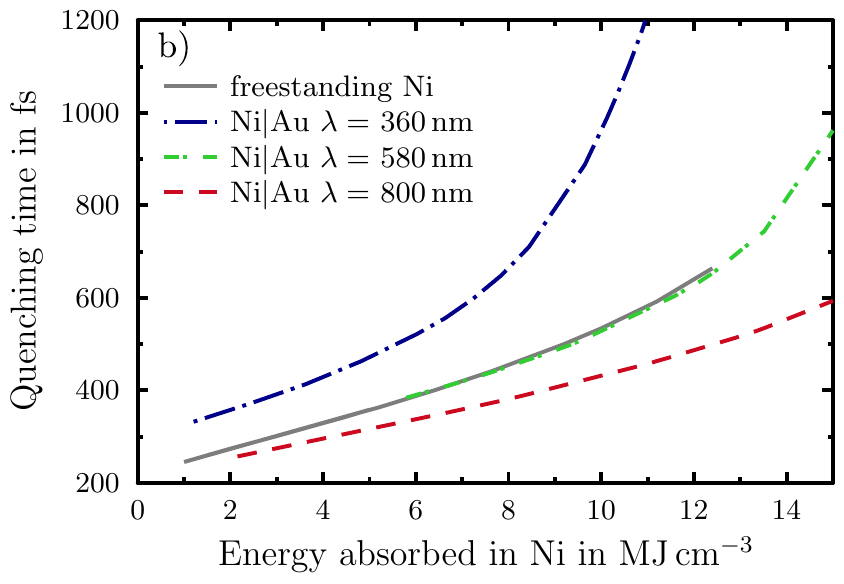}
    \caption{a) Comparison of calculated magnetization curves with same maximum quenching. The width of the trace around the minimum increases with decreasing wavelength. 
    b) Time of the minimum of the magnetization in dependence on the energy absorbed in Ni for different wavelengths. Longer wavelengths lead to a faster quenching.
    In both cases, the behavior of the bilayer depends significantly on the wavelengths.  \label{fig:magnetization_comparison}}
\end{figure}

\autoref{fig:magnetization_comparison} shows time-dependent magnetization traces of the Ni layer of the Ni/Au bilayer system calculated with the $\mu$TM. We selected magnetization traces for three characteristic excitation wavelengths and same maximum quenching.

At first glance, all magnetization traces show the typical lineshape known from  ultrafast demagnetization of ferromagnetic materials. They first reveal an ultrafast loss of magnetic order within the first hundreds of fs followed by a two-step remagnetization process with a fast, few ps, and a subsequent slower remagnetization time of up to several ten ps.
However, these 
curves also exhibit clear 
systematic changes when altering the wavelength of the optical excitation. The most distinct observation is the increasing quenching time (the time of the minimum of the magnetization, i.e. maximum quenching) with decreasing excitation wavelength. This coincides with a larger width of the magnetization traces around their magnetization minimum for shorter wavelength. In other words, the suppression of the magnetic order persists for a longer time when decreasing the wavelength of the excitation.

These characteristic 
differences
can be directly linked to the temperature gradients within the magnetic/non-magnetic bilayer system as has been shown in \autoref{fig:temperature_difference}. 
To this end we correlate 
the quenching time to the energy absorbed in Ni, see \autoref{fig:magnetization_comparison}~b). 
The quenching time for a negligible temperature gradient between Ni and Au (excitation with  \SI{580}{\nm} pulses) 
shows the same behavior as for a freestanding Ni layer of same thickness.
This hence reflects the intrinsic quenching time of the Ni layer without significant energy exchange with its environment. 
In contrast, an energy transfer from Ni to Au 
(as induced with \SI{800}{\nm} excitation)
decreases the quenching time  
for all absorbed energies, while an energy transfer from Au into Ni 
(as caused by \SI{360}{\nm} excitation)
increases the quenching time. 
In the latter case, the Au layer serves as an energy bath that successively provides energy for the demagnetization process of the Ni layer \cite{Fognini2015}. Crucially, the difference between the quenching time for the bilayer system with and without energy transfer increases with increasing magnitude of the temperature gradient between Ni and Au. In this way, we can uncover the quenching time as a characteristic signature of the magnitude and the sign of the energy transport between different layers of a magnetic/non-magnetic bilayer system. 

\begin{figure}
    \centering
    \includegraphics[width=\columnwidth]{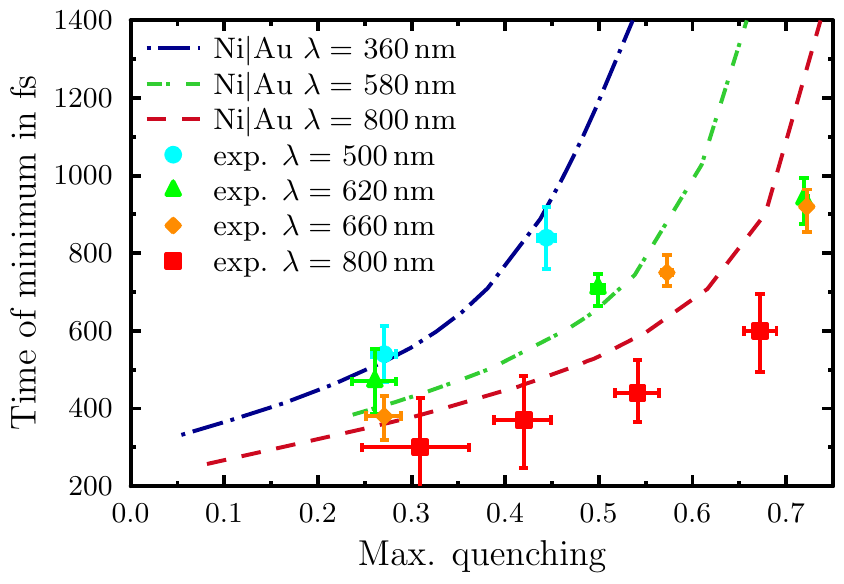}
    \caption{Time of the minimum in dependence on the maximum quenching for different wavelengths together with experimental data obtained from C-MOKE. Experiment and theory show the same order of magnitude and trend of wavelength ordering. 
    \label{fig:wavelength_comparison}}
\end{figure}

Finally, we demonstrate the predictive power of our model 
by comparing our simulated magnetization traces with experimental data of a Ni/Au bilayer structure for different excitation wavelengths. The magnetization dynamics is monitored experimentally by the all-optical \textit{C-MOKE} technique. \textit{C-MOKE} is a magneto-optical Kerr technique that allows us to determine the layer-specific magnetization dynamics of magnetic/non-magnetic bilayer systems~\cite{Hofherr2017,Hamrle2002,Schellekens2014}. In particular, we focus again on the characteristic signature of the energy transport between the Ni and Au layer. To this end, 
\autoref{fig:wavelength_comparison} shows the experimentally determined quenching time extracted for four different excitation wavelengths of \SI{500}{\nm}, \SI{620}{\nm}, \SI{660}{\nm}, and \SI{800}{\nm} depending on the maximum suppression of the magnetization.
These values are superimposed onto the simulated quenching-time vs. maximum-quenching traces for the three characteristic wavelengths discussed above. 
While smaller quantitative differences between the experimental and simulated data originate mainly from uncertainties of parameters used in the model, the experimental findings overall agree with our theoretical results.
They confirm the
theoretical prediction of a larger quenching time for shorter wavelengths and larger maximum quenching. 


In conclusion, our comprehensive extension of the \mutml has uncovered a strong and systematic variation of the sign and magnitude of the energy transfer in magnetic/non-magnetic bilayer systems that depends on the wavelength of the optical excitation. For the particular case of the Ni/Au bilayer structure, optical excitation with small wavelength in the UV range leads to an energy transfer from Au to Ni while the direction of the energy transfer is reversed for excitation with large wavelength in the IR range. Our findings hence clearly demonstrate the potential to shape temperature gradients in multilayer stacks by modulation of the excitation wavelength. This opens new opportunities to optically control, for instance, the energy dissipation efficiencies or the suppression time of the magnetic order of individual layers in magnetic multilayer structures. Thus, it allows and hence to steer and control spin and charge carrier functionalities in the next generation of spintronic assemblies. 

This work was funded by the Deutsche Forschungsgemeinschaft (DFG, German Research Foundation) - TRR 173 - 268565370 Spin+X (Projects A08 and B03). C.S. acknowledges fruitful collaboration with Markus Uehlein in developing the simulation software, B.S. acknowledges financial support by the Dynamics and Topology Center funded by the State of Rhineland Palatinate.

    \bibliography{bibfile/all.bib}
    
\end{document}


\title{Supplementary Information: Control of transport phenomena in magnetic heterostructures by wavelength modulation}
    \author{Christopher~\surname{Seibel}}
    \email{cseibel@physik.uni-kl.de}
    \author{Marius~Weber}
    \author{Martin~Stiehl}
    \author{Sebastian~T.~Weber}
    \author{Martin~Aeschlimann}
    \author{Hans~Christian~Schneider}
    \affiliation{Department of Physics and Research Center OPTIMAS, Technische Universitaet Kaiserslautern, 67663 Kaiserslautern, Germany}
    \author{Benjamin~Stadtmüller}
    \affiliation{Department of Physics and Research Center OPTIMAS, Technische Universitaet Kaiserslautern, 67663 Kaiserslautern, Germany}
    \affiliation{Institute of Physics, Johannes Gutenberg University Mainz, Staudingerweg 7, 55128 Mainz, Germany}
    \author{Baerbel~Rethfeld}
    \affiliation{Department of Physics and Research Center OPTIMAS, Technische Universitaet Kaiserslautern, 67663 Kaiserslautern, Germany}

    \date{\today}
    \maketitle
\section{The \mutm for one material}

For a homogeneously heated magnetic material, the \mutm describes the changes of the 
internal energy $u$ for the spin-resolved electronic bands and the phonons, respectively, as well as the change of the particle densities $n$ within the electronic bands:
\begin{subequations} \label{eq:mutm_rhs}
\begin{align}
	\dv{u_{e}^\sigma}{t}   &= -\gamma( T^\sigma - T^{\overline{\sigma}}) - g^\sigma(T^\sigma - T_p) + s^\sigma(t), \label{eq:muTM_spin_dudt}\\
	\dv{u_p}{t}            &= -g^\uparrow(T_p - T^\uparrow) - g^\downarrow(T_p - T^\downarrow),\label{eq:muTM_spin_dudt_ph}\\
    \dv{n_e^\sigma}{t}   &= -\nu (\mu^\sigma - \mu^{\overline{\sigma}}), \label{eq:muTM_spin_dndt}
\end{align}
\end{subequations}
where the index $e$ denotes electrons and $p$ the phonons, respectively. The superscript $\sigma$ stands for the spin direction, $\sigma \in \{\uparrow, \downarrow \}$, the opposite direction
is given by 
$\overline{\sigma}$.

The energy change due to temperature difference is determined by the electron-phonon coupling parameter $g$. 
Similarly, the rate of the energy exchange between the electronic subbands is governed by the coupling parameter $\gamma$.
The equilibration of chemical potentials leads to changes of the particle densities through the coupling parameter $\nu$.
The considered processes are sketched for the case of nickel in the grey-shaded area of Figure 1 of the main text. 
Solid lines refer to an energy transfer based on the equilibration of temperatures and dashed lines refer to a particle transfer based on the equilibration of chemical potentials.

Since energy and particle density depend on the temperature as well as on the chemical potential, the temporal change of the former can be translated into changes of the latter as described in Ref.~\cite{Mueller2014PRB}.
The particle and energy densities are given by the 0th and 1st moment of the distribution function, uniquely determined by the time- and spin-dependent chemical potentials and temperatures.
Here, we assume the same DOS for up- and down-electrons, shifted by the exchange splitting $\Delta$ 
given as $\Delta = m \, V_0\, U$, where $V_0$ is the unit cell volume per atom, $U$ the Stoner parameter (cf. \autoref{tab:material_parameters})~\cite{Mueller2013PRL}, and $m= n_e^\uparrow - n_e^\downarrow$ the transient magnetization.

\section{Parameters of the \mutm for two layers}

In \autoref{tab:material_parameters}, the constant material-specific parameters of Ni and Au are listed, which were used for the calculations. Although it is known that the electron-phonon coupling parameter depends, e.g., on the electron and phonon temperatures and the electron density~\cite{Lin2008, Mueller2013PRB, Medvedev2020}, we assume it to be constant as usually done for temperature-based models~\cite{Pudell2018,Anisimov1974,Mueller2014PRB,Sokolowski-Tinten2017}. 
We use the values at \SI{1000}{\kelvin}, because this corresponds approximately to the center of the range of reached electron temperatures. 
\begin{table}
    \centering
    \caption{Material-specific parameters of Ni and Au used for the calculations.\label{tab:material_parameters}}
    \begin{tabular}{>{\RaggedRight}p{0.65\columnwidth}p{0.15\columnwidth}p{0.15\columnwidth}}
        \toprule
        Parameter & Nickel & Gold\\
        \midrule
        Lattice specific heat, $c_{ph}$ (\SI{e6}{\joule\per\meter\cubed\per\kelvin}) & \num{3.8}& \num{2.5}\\
        Electron-phonon coupling constant @\SI{1000}{\kelvin}, $g$ (\SI{e17}{\watt\per\meter\cubed\per\kelvin}) & \num{24.0}\cite{Mueller2013PRB} & \num{0.26}\cite{Lin2008, Mueller2013PRB} \\
        Electron-electron temperature coupling, $\gamma$ (\SI{e17}{\watt\per\meter\cubed\per\kelvin}) & \num{500} & \num{5.4} \\
        Electron-electron chemical potential coupling, $\nu$ (\SI{e60}{\per\joule\per\meter\cubed\per\second}) & \num{5.0} & \num{5.0} \\
        Stoner parameter, $U$ (\SI{}{\electronvolt}) & \num{0.506} & \num{0.2}\cite{Sigalas1994}\\
        \bottomrule
    \end{tabular}
\end{table}

\begin{table}
    \centering
    \caption{Model parameters responsible for the energy and particle transfer at the interface.\label{tab:interface_parameters}}
    \begin{tabular}{>{\RaggedRight}p{0.65\columnwidth}p{0.15\columnwidth}p{0.15\columnwidth}}
        \toprule
        Parameter & Nickel & Gold\\
        \midrule
        Seebeck coefficient, $\Sigma$ (\SI{e-6}{\volt\per\kelvin})& -20.0\cite{Lasance2006} & 1.5\cite{Lasance2006}\\
        Peltier coefficient @\SI{500}{\kelvin}, $\Pi$ (\SI{e-3}{\volt}) &-10.0 &0.75\\
        Thermal boundary conductance, $\sigma_{th}$ (\SI{e9}{\watt\per\meter\squared\per\kelvin})	& \multicolumn{2}{c}{2.00}\\
        Specific contact conductance, $\sigma_c$ (\SI{e11}{\per\ohm\per\meter\squared}) & \multicolumn{2}{c}{1.00}\\
        $\kappa_{u,T}$ (\SI{e9}{\watt\per\meter\squared\kelvin}) & \multicolumn{2}{c}{2.00}\\
        $\kappa_{u,\mu}$ (\SI{e27}{\per\meter\squared\per\second}) & \multicolumn{2}{c}{2.89}\\
        $\kappa_{n,T}$ (\SI{e24}{\per\meter\squared\per\kelvin\per\second}) & \multicolumn{2}{c}{5.77}\\
        $\kappa_{n,\mu}$ (\SI{e48}{\per\meter\squared\per\joule\per\second}) & \multicolumn{2}{c}{3.90}\\
        Spin-up interface reflectivity, $R^\uparrow$ & 0.23\cite{Lu2020} & 0.19\cite{Lu2020}\\
        Spin-down interface reflectivity, $R^\downarrow$ & 0.75\cite{Lu2020}& 0.26\cite{Lu2020}\\
        \bottomrule
    \end{tabular}
\end{table}

To derive the parameters of the energy and particle transfer across the interface, we use the general approach of temperature-based model. We assume the energy transfer to be proportional to the temperature difference and the particle transfer to be proportional to the difference of the chemical potentials. 
We assume the coupling constant of the energy transfer due to the difference of temperatures to equal the thermal boundary conductance,
\begin{equation}
    \kappa_{u,T} = \sigma_{th}\enspace.
\end{equation}
This assumption is very common in modeling metal-dielectric interfaces~\cite{Sokolowski-Tinten2017, Hopkins2007}. We estimate the order of magnitude of the thermal boundary conductance from time-domain thermoreflectance measurements~\cite{Gundrum2005}.

Similarly, the coupling parameter for the particle transfer due to the difference of the chemical potentials is modeled with the specific contact conductance:
\begin{equation}
    \kappa_{n,\mu} = \frac{\sigma_{c}}{e^2}\enspace,
\end{equation}
where $e$ is the elementary charge. 

Additionally, the difference of the chemical potentials leads to an energy transfer, which is described by the Peltier effect. From that, the respective coupling parameter is
\begin{equation}
    \kappa_{u,\mu} = - \frac{\Pi\sigma_{c}}{\abs{e}}\enspace,
\end{equation}
with $\Pi$ being the Peltier coefficient. The inverse thermoelectric effect, the Seebeck effect, describes the particle flow due to a temperature gradient. Then, the respective coupling parameter can be derived to be
\begin{equation}
    \kappa_{n,T} = - \frac{\Sigma\sigma_{c}}{\abs{e}}\enspace,
\end{equation}
where $\Sigma$ is the Seebeck coefficient, which is related to the Peltier coefficient via $\Pi = \Sigma T$. For simplicity and to ensure particle and energy conservation, we assume the average of the thermoelectric coefficients of both materials at a constant temperature of \SI{500}{\kelvin}, which resembles the equilibrium temperature of our system after the laser excitation. 
The resulting parameters are given in \autoref{tab:interface_parameters}.

\section{Spin-dependent absorption}

For our results, we assume an equal absorption of up and down electrons, i.e. we distribute the absorbed energy equally to the up and down bands. Here we show that this assumption is justified.
\autoref{fig:up_down_absorption} compares magnetization curves for our Ni|Au bilayer calculated for three different edge cases of spin-dependent absorption: 
\begin{figure}[H]
    \centering
    \includegraphics[width=\columnwidth]{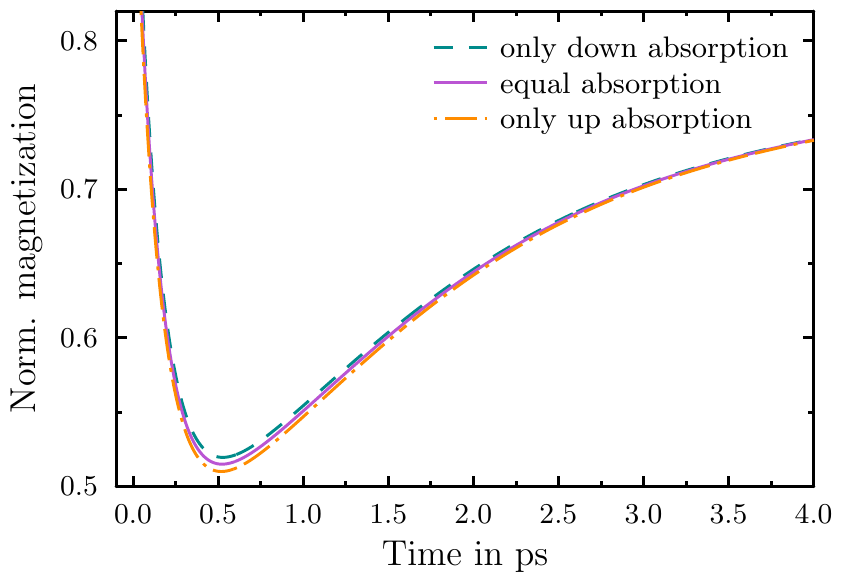}
    \caption{Comparison of magnetization curves for various ratios of up and down absorption. \label{fig:up_down_absorption}}
\end{figure}
(i) the down electrons absorb all the energy, so the up electrons do not absorb, 
(ii) the absorbed energy is distributed equally to both bands as used in the main text, and (iii) the up electrons absorb all the energy, so the down electrons do not absorb.
The laser fluence was chosen in way that the reached quenching is around \num{0.5} to be comparable to most of our calculations. 

The three curves exhibt no differences in the demagnetization and in the remagnetization after around \SI{3}{\ps}. Only around the minimum of the magnetization tiny deviations are presented. However, these are on the order of about \SI{1}{\percent}, which is well below the experimental resolution and the magnitude of the effect we present in the main text. Thus, in the framework of our model it is indeed justified to assume an equal absorption of up and down electrons for simplicity. 

\section{Optical Absorption }
We calculate the electric field in the multilayer structure by solving
the Helmholtz equation for the complex field amplitude $E (\boldsymbol{z})$ by a finite-difference method. As boundary condition we impose an incoming wave from the left and obtain the depth-dependent absorption from~\cite{Khorsand2014} 
\begin{equation}
    \mathrm{d}A(z)=\alpha(z) \tilde{n}(z) |E(\boldsymbol{z})|^2 \mathrm{d}z .
\end{equation}
 The refractive index is here denoted by $n=\tilde{n}+ik $ with real part $\tilde{n}$ and imaginary part ${k}$ and $\alpha=4 \pi k/\lambda $. We use \textit{ab initio} values for the wavelength ($\lambda$) dependent refractive indices~\cite{Werner2009} and assume normal incidence. 
    \bibliography{bibfile/all.bib}